\documentstyle[11pt,newpasp,twoside]{article}
\def\edcomment#1{\iffalse\marginpar{\raggedright\sl#1\/}\else\relax\fi}
\reversemarginpar

\begin{document}
\title{On The Origin Of Subdwarf B Stars and Related Metal-Rich Binaries}
\author{Elizabeth M.\ Green \& James Liebert}
\affil{Steward Observatory, University of Arizona, Tucson, AZ 85721}
\author{Rex A.\ Saffer}
\affil{Dept.\ of Astronomy \& Astrophysics, Villanova University, 
  Villanova, PA 19085}

\begin{abstract}
Mounting evidence from subdwarf B (sdB) stars in the galactic field
and their recently discovered counterparts in old open clusters
indicates that at least two thirds of local disk sdB stars are
binaries.  Our recent radial velocity survey showed that sdB binaries
naturally divide into two groups with contrasting spectroscopic and
kinematic properties.  Those with detectable spectral lines from a
cooler companion invariably have periods longer than a year, while
very short period sdB's have essentially invisible companions.  We
derive typical orbital separations for the components of the composite
spectrum sdB's from their velocities.  The current systems must have
been produced by Roche lobe overflow/mass transfer from low mass,
metal-rich giants near the first red giant branch tip, without
undergoing common envelope envolution.  The same process should also
occur at slightly lower red giant luminosities, producing a wide
binary with a helium white dwarf instead of an sdB star.  Most short
period sdB's probably result from a common envelope following Roche
lobe overflow of the initial secondary onto the white dwarf.  Rare
post-common envelope sdB + main sequence (MS) binaries also exist, but
available data suggest that most such systems involving lower MS
companions end up merging.  The small and nearly identical masses of
the two known MS survivors in short period sdB binaries imply that
both components must have lost a large fraction of their initial mass
in the common envelope process.
\end{abstract}

\section{Introduction}

Field subdwarf B (sdB) stars are core helium burning objects commonly
found in the galactic disk, and also recently identified in two old
open clusters (Liebert, Saffer, \& Green 1994).  They are believed to
be extreme horizontal branch stars with $M_{\rm sdB} = 0.5 M_{\sun}$
({\it e.g.}\ Saffer et al.\ 1994), produced when the envelope is
stripped from a low mass giant near the first red giant branch tip.
Whether this is due primarily to mass transfer followed by common
envelope ejection in binaries (Mengel, Norris, \& Gross 1976) or
enhanced mass loss in single stars (D'Cruz et al.\ 1996) is still not
clear.  The mechanisms for the extreme mass loss required to create
significant numbers of sdB's and low mass white dwarfs in low density
environments are important for understanding the evolution of
many astrophysically interesting systems, including millisecond
pulsars, low-mass X-ray binaries, double degenerate white dwarfs, and
possibly the origin of type Ia supernovae, yet they remain very poorly
understood.  SdB's are also the main candidate for the unexpected
far-UV excesses seen in many external galaxies (O'Connell 1999, and
references therein).  They would be a valuable probe of ages and
metallicities if their evolutionary behavior could be uniquely
modeled.  Most recent investigators consider sdB stars a result of
single star evolution ({\it e.g.}\ Bressan, Chiosi, \& Fagotto 1994;
Dorman, O'Connell, \& Rood 1995; Park \& Lee 1997; Yi, Demarque, \&
Oemler 1998; Yong, Demarque, \& Yi 2000).

Meanwhile, studies of local field sdB's provide increasing evidence
for a large binary fraction, initially including estimates from
broadband flux distributions (Allard et al.\ 1994), more recently from
growing numbers of newly discovered short period sdB's (Menzies 1986;
Kilkenny et al.\ 1998; Koen, Orosz, \& Wade 1998; Moran et al.\ 1999;
Bill\`eres et al.\ 1999; Maxted et al.\ 2000).  Our radial velocity
survey for a large sample of bright field subdwarf B stars (Saffer,
Green, \& Bowers, this volume) has now demonstrated that at least two
thirds of disk sdB's are in binaries.  Some 20\% show spectral lines
from a cool turnoff or subgiant companion, and have relatively long
periods of at least a year.  Another 45\% of sdB's are clearly
post-common envelope binaries with periods of hours or days, whose
companions are always too faint to be detected spectroscopically.

Using our observed data for field sdB's and their probable progenitors
in old open clusters, we propose two related scenarios to explain the
evolution of both types of sdB binaries, and discuss the implications
for common envelopes.

\section{Discussion}

\subsection{Clues from Old Open Clusters}

The only open clusters known to contain sdB stars, NGC\ 6791 and NGC\
188, are also the metal-richest ([Fe/H] = +0.4 and $-$0.05,
respectively) of the very old open clusters.  Thus, their stellar
properties include both lower turnoff masses ($1.1 - 1.25 M_{\sun}$,
Chaboyer, Green, \& Liebert 1999; Liu \& Chaboyer 2000) and larger
giant radii ($150~ \lower 0.5ex\hbox{$\buildrel < \over \sim\ $} R\
\lower 0.5ex\hbox{$\buildrel < \over \sim\ $} 200$ at the first red
giant tip, Salasnich et al.\ 2000) than most disk stars.  NGC\ 6791
contains two of the three cataclysmic variables so far found in open
clusters (Ka\l u\.zny et al.\ 1997).  Curiously, both clusters also
show an enormous fraction of blue stragglers, N$_{\rm
blue~straggler}$/N$_{\rm horizontal~branch}\ \lower 0.5ex\hbox{$\buildrel 
> \over \sim\ $} 3$ (Green 2000, in preparation).
In M80, a postulated pre-core collapse cluster with the highest known
blue straggler frequency of any globular, the same ratio is only about
1 (Ferraro et al.\ 1999).

\subsection{The Composite Spectrum SdB Stars are the Key}

The velocity differences between the two components of the composite
spectrum sdB binaries can be used to estimate typical values for their
orbital periods and separations.  The mean $|\Delta v \sin i|$ from 89
observations of 19 composite binaries over a two year period was about
11.5~km~s$^{-1}$ (the largest difference was 30~km~s$^{-1}$).
Assuming random orbital inclinations and typical old disk
turnoff/subgiant masses of $1.0 - 1.3 M_{\sun}$ for the companions
(required to match the observed luminosity contributions), the current
periods average $3 - 4$ years with separations of $540 - 650
R_{\sun}$.  If the sdB's evolved without interacting with their
companions, their prior separations can be estimated by assuming
$a(M_1 + M_2)$ = constant (Tout et al.\ 1997).  With initial sdB
progenitor masses similar to the turnoff masses in NGC\ 6791 and NGC\
188, and mass loss of $\sim 0.3 M_{\sun}$ on the first giant branch,
typical separations just prior to the He core flash would have been
$415 - 520 R_{\sun}$.  The corresponding effective radii of the
giants' Roche lobes (Eggleton 1983) would have been $155 - 185
R_{\sun}$.  These are of the order of, and in most cases slightly
smaller than, the required radii at the red giant tip!

The inescapable conclusion is that composite spectrum sdB's were
stripped of their envelopes due to Roche lobe overflow of their
progenitor giants just before the first red giant tip, without
developing a common envelope.  Therefore, despite canonical
expectations (Iben \& Livio 1993), nearly stable Roche lobe overflow
on the upper giant branch must occur relatively often, given the right
circumstances.  We suggest that sufficiently low mass ($\lower
0.5ex\hbox{$\buildrel < \over \sim\ $} 1.3 M_{\sun}$), preferentially
metal-rich ([Fe/H] $\lower 0.5ex\hbox{$\buildrel > \over \sim\ $}
0.0$) red giants in suitably wide binaries may transfer enough mass to
initially massive secondaries during the giant's normal slow mass loss
phase, to reduce the mass ratio close to the critical value required
for stable mass transfer.  Complete stability would not be required
initially, as long as the near-solar mass secondary is able to accept
the dynamic mass transfer of the first couple of tenths of a solar mass
without filling its giant size Roche lobe ($\sim 170 R_{\sun}$ and
increasing).

\subsection{Further Speculations}

When the primary in the above scenario attains a large enough core
mass prior to Roche lobe overflow to enable eventual He-burning, the
result is a composite spectrum sdB with a probable blue straggler
companion in a widened orbit.  Overflow at slightly lower luminosities
would produce a He white dwarf instead of an sdB, everything else
remaining the same.  We propose that a significant fraction of the
blue straggler+white dwarfs become post-common envelope sdB+wd
binaries following Roche lobe overflow of the initial secondary, since
the observed number of short period sdB binaries is more than twice
the number of composite sdB's.  Populations that produce these sdB
stars require fairly low densities for their lengthy evolution and
will contain many long period blue stragglers.

A common envelope will occur if the first Roche lobe overflow
begins after the primary develops a deep convective envelope but
before sufficient mass is lost/transferred to the secondary, or if the
initial mass of the secondary is too small.  Both cases would be
assumed to produce numerous low mass cataclysmic or pre-cataclysmic
variables.  However, any reasonable initial mass function ought to create
many more post-common envelope sdB+MS binaries than long period
composite spectrum sdB's, since only the most massive secondaries can
avoid a common envelope.  Yet, we note that only two short period
sdB+MS binaries have been found, compared to at least four sdB+wd
systems (and counting), despite extensive photometric monitoring of
several hundred sdB's (Koen et al.\ 1998).  This implies a genuine
scarcity of post-common envelope sdB+MS binaries, which are much more
easily detected by reflection effects and eclipses.  (MS stars with
masses $\lower 0.5ex\hbox{$\buildrel > \over \sim\ $} 0.14 M_{\sun}$
have larger radii than sdB stars.)  Thus, most MS companions in common
envelope systems with initial mass ratios, $M_2 / M_1$, sufficiently
below unity might end up merging.

The $\sim 0.14 M_{\sun}$ mass secondaries in the two known eclipsing
sdB+MS systems would then represent the surviving remnants from a very
narrow initial mass range: too small to escape a common envelope, but
large enough to avoid merging.  If so, they must have lost a large
fraction of their original mass along with the sdB progenitor in the
common envelope process.

\vfill
\end{document}